\documentstyle[prl,aps,floats,twocolumn,epsf]{revtex}
\begin{document}                % INITIALIZE - DONT CHANGE
\textheight=9.45in
\twocolumn[\hsize\textwidth\columnwidth\hsize\csname @twocolumnfalse\endcsname
\title{Missing $2k_{F}$ Response for Composite Fermions in Phonon Drag}
\author{S. Zelakiewicz, H. Noh, T. J. Gramila}
\address{Department of Physics, The Pennsylvania State University, 
University Park, Pennsylvania 16802}
\author{L. N. Pfeiffer, K. W. West}
\address{Bell Labs, Lucent Technologies, Murray Hill, New Jersey 07974}

\maketitle
\begin{abstract}                % DON'T CHANGE THIS LINE

The response of composite Fermions to large wavevector scattering has
been studied through phonon drag measurements.  While the response
retains qualitative features of the electron system at zero magnetic
field, notable discrepancies develop as the system is varied from a
half-filled Landau level by changing density or field.  These
deviations, which appear to be inconsistent with the current picture of
composite Fermions, are absent if half-filling is maintained while
changing density.  There remains, however, a clear deviation from the
temperature dependence anticipated for $2k_F$ scattering.

\end{abstract}
]

Composite Fermions (CF), new quasiparticles initially described as the
combination of an electron with an even number of magnetic flux
quanta, provide a simplifying physical picture of the fractional
quantum Hall effect (FQHE)\cite{jain:cf}.  The particles have also
been argued\cite{hlr:cf} to possess many of the properties of
electrons at zero magnetic field, experiencing an effective field
which is zero for a half-filled Landau level even though they exist in
the presence of extreme magnetic fields.  Numerous experimental
investigations, including studies of surface acoustic waves
(SAW)\cite{willett:saw}, cyclotron resonance in antidot
lattices\cite{stormer:antidot}, activation energies\cite{du:gaps} and
magnetic focusing\cite{goldman:mf}, have confirmed the existence of
these particles and reveal behavior similar to zero-field electrons.
The experiments, in addition, clearly support the existence of a Fermi
surface for the particles.  A common element of these investigations,
however, is that they have generally been limited to small wavevector
scattering.  A key question for the particles, how they respond to
large wavevector scattering, especially across the re-emergent Fermi
surface, has not been systematically investigated in experiment.  It
is this response that the experiments presented here were designed to
address.

Access to the large wavevectors required for scattering a CF across
the Fermi surface is provided here through phonons.  The use of
phonons permits the scattering wavevector (q) to be effectively
tunable by changing the temperature, T.  At low temperatures, only
small wavevector acoustic phonons are thermally excited, limiting
scattering of CF to small q processes.  As the temperature is
increased, access to larger wavevector phonons permits larger
wavevector scattering.  For electrons, the transition to large angle
scattering is readily evident due to a sharp cutoff\cite{dynes:2kf}
for scattering with q greater than twice the Fermi wavevector
(2k$_F$).  The cutoff results directly from the existence of a Fermi
surface and the combined restrictions of momentum and energy
conservation.  A clear change in temperature dependence, the
Bloch-Gr\"uneisen transition, has been directly observed in
resistivity measurements\cite{stormer:bg} and in phonon
drag\cite{gramila:vphonon}.

A similar cutoff should exist for composite Fermions.  Two key
elements are required.  The first is the presence of a Fermi surface
for the particles.  The second is that CF must be able to withstand
large wavevector scattering, sending a CF across its Fermi surface.
While the first condition is well
established\cite{willett:saw,stormer:antidot,goldman:mf}, the second
has not been theoretically investigated in detail, with studies
generally limited to small wavevectors and low temperatures.  Details
of the phonon interaction with CF should not affect the existence
of this cutoff, as they do not for electrons, which depends only on
the magnitude of the phonon wavevector as compared to 2k$_F$.

The isolation of phonon scattering in this work is attained through
electron drag measurements between remotely spaced parallel
two-dimensional electron gas (2DEG) layers.  In electron
drag\cite{gramila:drag} when a current is driven through one of two
electrically isolated 2DEG's, interlayer electron-electron (e-e)
interactions transfer momentum to the second layer, inducing a voltage
in that layer.  The drag transresistivity $\rho_D$, the ratio of this
voltage to the applied current per square, is a direct measure of the
interlayer scattering rate\cite{gramila:drag}.  While Coulomb
scattering dominates $\rho_D$ for closely spaced layers, its strong
layer spacing
dependence\cite{gramila:drag,theory:drag}
permits interlayer phonon exchange to completely
dominate\cite{gramila:vphonon,noh:phonon} interactions of
remote layers.

The samples used in this work, GaAs/AlGaAs double quantum well
structures, consist of two 200\AA\ wide quantum wells which are
remotely spaced.  The bulk of the CF measurements were performed on a
sample with a 5000~\AA\ barrier thickness.  Each layer has an electron
density, n, near $1.5\times10^{11}/cm^{2}$ as grown, with mobilities
approaching $2\times10^{6} cm^{2}/Vs$.  Individual layer densities
were varied through the application of a voltage to an overall top
gate or by applying an interlayer bias, with the densities in each
layer made equal for all measurements.  The large two-terminal
resistances present in the samples at high fields demanded particular
care, requiring measurement frequencies as low as 0.5 Hz and currents
as low as 20 nA.  Established tests\cite{gramila:drag} such as
interchanging current and voltage leads, testing current linearity,
and ensuring the absence of interlayer leakage and other spurious
signals through ground configuration changes all confirm the validity
of the measurements.  The lack of change in the drag signal upon
reversal of the magnetic field indicates that Hall voltages play no
role in these measurements.  Comparable results were obtained for a
second 5000 \AA\ barrier sample and a 2400 \AA\ barrier sample.

The effect of the 2k$_F$ cutoff for phonon scattering is shown in
Fig.\ 1a for a zero-field phonon drag measurement on a 2400 \AA\
barrier sample in which Coulomb scattering is negligible.  Data are
plotted as $\rho_D/T^2$, revealing a distinct change in temperature
dependence and a peak near 2 K.  The peak position is known not to
change with layer spacing\cite{gramila:vphonon,noh:phonon}; $\rho_D$
for this sample is shown due to significantly reduced signals for the
5000~\AA\ barrier sample at zero field.  The peak position, which
varies with the size of the Fermi surface (i.e.,
$\propto\sqrt{n}$)\cite{noh:phonon,all:phdrag}, quantifies the
transition from a strong temperature dependence with q $\leq$ 2k$_F$,
to the weaker dependence when q is limited to 2k$_F$ scattering.  The
inset plots the relative net momentum carried by phonons of a given
in-plane wavevector for both deformation potential and piezo-electric
coupling at 3 K.  This single-layer calculation, based closely on
earlier work\cite{stormer:bg,gramila:vphonon}, clearly shows the
cutoff at 2k$_F$ is independent of details of the electron-phonon
interaction.  The temperature of the peak in $\rho_D/T^2$ is thus directly
related to the phonon wavevector which matches 2k$_F$.

Before examining the temperature dependence of phonon drag for
composite Fermions, it is necessary to re-establish, at high fields,
that phonon scattering dominates $\rho_D$.  This is explored through
measurements made below 1K (Fig.\ 1b, inset) on a 5000 \AA\ barrier
sample at 13T corresponding to a half-filled lowest Landau level ($\nu
= 1/2$).  This data is well characterized by a power law
dependence with a best fit of $\rho_{D}\propto T^{3.7}$ (solid line).
This exponent is substantially higher than the sub-quadratic
dependence established for Coulomb drag of
CF\cite{lilly:cfdrag,stern:cfdrag,kim:cfdrag,sakhi:cfdrag} and is more
consistent with expectations of phonon scattering from thermopower
measurements\cite{maan:cftp} and theoretical
calculations\cite{reizer:cfphonon}.  The behavior of $\rho_D$ at
low temperatures firmly establishes a negligible role for Coulomb
scattering in this sample.

\begin{figure}[tb]
\begin{center}
\leavevmode
\hbox{%
\epsfxsize=3in
\epsffile{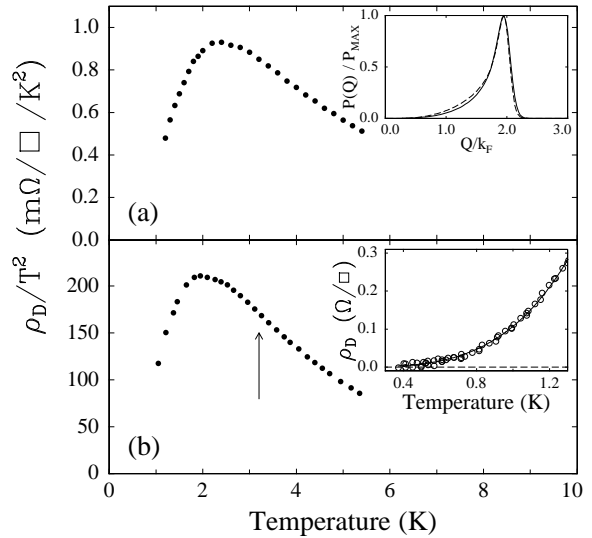}}
\end{center}
\caption{ T dependence of $\rho_{D}/T^2$, with density $1.5\times
10^{11}$, shows similar behavior for (a) zero-field electrons and (b)
composite Fermions.  Arrow indicates the anticipated peak position for
CF.  Upper inset: Relative net momentum transfer rate at 3 K for
single layer phonon scattering vs.\ q, showing 2k$_F$ cutoff for
deformation potential (solid line) and piezo-electric (dashed line)
coupling.  Lower inset: Low temperature $\rho_{D}$ for composite
Fermions.  A fit proportional to $T^{3.7}$ indicates the absence
of Coulomb scattering.}
\label{1}
\end{figure}

Measurements of $\rho_D/T^2$ for CF at higher temperatures, shown in
Fig.\ 1b, reveal a behavior remarkably similar to that for zero field
electrons at the same density.  The transition from a strong to a weak
temperature dependence mimics the low field data, with a peak position
near but slightly lower than that in Fig.\ 1a.  The behavior indicates
a distinct wavevector cutoff in the phonon scattering process.  A
notable difference is the magnitude of $\rho_D$, being significantly
larger for CF.  This increase is similar to the enhanced scattering of
CF generally observed.

While the data confirm the existence of a wavevector cutoff, the
temperature of the peak in $\rho_D/T^2$, $T_P$, is substantially lower
than expected.  Spin polarization of the CFs results in a larger Fermi
surface than at zero field, yielding a peak in $\rho_D/T^2$ at a higher
temperature for the same phonon system.  The expected $\sqrt{2}$
increase in the size of the Fermi surface has been established in
other measurements\cite{willett:saw,stormer:antidot,goldman:mf} and
would result in a peak position closer to 3 K as indicated by the
arrow in the figure.

The substantial difference between the measured and anticipated peak 
position for CF raises the possibility that the q cutoff may not result
from the CF Fermi surface.  Questions of CF stability, for example,
must be considered.  Theoretical predictions\cite{morf:cfstability} of
the CF binding energy are $\sim 4K$ for these densities.  
The observed peak position, 1.9 K, however, is below this binding
energy and well within the range for which CF effects are observable
in SAW measurements\cite{willett:saw2}.  The lack of strong FQHE
states at these temperatures does not indicate an invalid regime for
CFs, but merely the absence of an energy gap.  This distinction is
evident in recent magnetization
measurements\cite{heitmann:magnetization}.

Another possibility is that the maximum in $\rho_D/T^2$ is due to
single-particle effects of the electron system in a high magnetic
field.  For example, the scattering wavevector may have a cutoff
determined by the width of the Landau level\cite{cfthermo} or the
magnetic length\cite{dietzel:hfphonon}.  These origins of a cutoff
have been argued to be responsible for features observed in earlier
thermopower measurements at $\nu=1/2$\cite{cfthermo-exper} and
ballistic phonon absorption at high magnetic
fields\cite{dietzel:hfphonon}, respectively.  Both of these mechanisms
would result in an increase of the peak position as the field is
increased.  However, examination of $\nu=1/4$ (not shown), another CF
state, shows a temperature dependence similar to $\nu=1/2$ for a given
density, but with a $\sim10\%$ lower peak position.  This small
decrease in $T_P$, for a factor of 2 increase in field, clearly
contradicts scattering limitations due to the Landau level width or
the magnetic length.  In addition, the similarity between the peak
position for $\nu=1/2$ and $\nu=1/4$ supports the assertion that
composite Fermions are observed.

To explore the origin of the discrepancy in the peak position,
$\rho_D$ was measured in the presence of an effective magnetic field.
Figure 2a shows the effect of varying the system away from $\nu = 1/2$
by changing the magnetic field with a constant density.  A striking
element of these measurements is the change in the magnitude of
$\rho_D/T^2$, which increases by roughly threefold.  Another is the
variation of the peak position with field.  The value of $T_P$ has
been quantified through a fit in the vicinity of the maximum, with the
resultant peak values, shown in the inset, generally insensitive to
the functional form of the fit.  At fields near and above half
filling, $T_P$ is proportional to $\sqrt{B}$ (solid line), while $T_P$
falls below this dependence at lower fields.

\begin{figure}[bt]
\begin{center}
\leavevmode
\hbox{%
\epsfxsize=3in
\epsffile{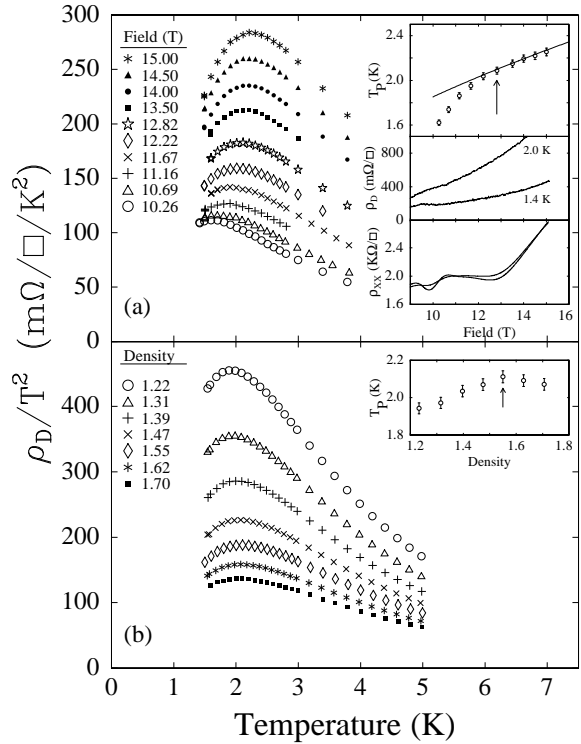}}
\end{center}
\caption{(a) T dependence of $\rho_{D}/T^2$ at various magnetic fields
with fixed density, $\nu=1/2$ occurs at 12.82T.  Upper inset: Peak
position vs.\ field.  Solid line represents $T_P \propto \sqrt{B}$ at
higher fields with the arrow indicating $\nu=1/2$.  Middle and lower
inset show $\rho_D$ and the longitudinal resistivity, at
1.4 K and 2.0 K, illustrating the vanishing effect of the fractional
states at these temperatures.
Plot (b) is
similar to (a) but with field fixed and density varied in units of ($
10^{11} cm^{-2}$).  These behaviors deviate from the anticipated
response of CF.  }
\label{2}
\end{figure}

A complimentary method for the application of an effective magnetic
field is explored in Fig.\ 2b, where the external field is constant
and the density is varied.  Significant changes in magnitude continue
to be present as the density is varied with the magnetic field fixed
at 12.8T.  Compared to the field dependence of Fig.\ 2a, however,
there is substantially less variation in the position of $T_P$
(inset), with a weak maximum at half filling.  This is suggestive that
half filling, and thus CF, are important in determining the cutoff.

The changes in magnitude and peak position with the application of an
effective field are inconsistent with general expectations for CF away
from half filling.  For example, field variations have been
observed\cite{stormer:antidot,goldman:mf} to induce cyclotron motion
of the composite particles, which experience an effective field equal
to the difference of the applied field from that at half filling.
Properties related to the Fermi surface of CF should persist for low
effective fields, as they do for bare electrons, until the period of
cyclotron motion is less than the scattering time.  From this
perspective, a peak position determined by the size of the Fermi
surface should not change over the range of effective fields explored
in Fig.\ 2 and the magnitude should remain relatively constant.  It is
thus difficult to reconcile the changes in the measured behavior
within a simple CF picture.

The complex behavior observed motivates consideration of spin effects,
though expectations of spin-splitting energies are large enough that
such effects appear unlikely.  Measurements of $\rho_D$, with the
sample tilted 22$^{\circ}$, matching the perpendicular fields and
electron density of Fig.\ 2a, were indistinguishable from that data in
both magnitude and peak position.  This rules out a role for spin in
the interlayer phonon scattering process.

A common element of the measurements of Fig.\ 2 is that significant
deviations from $\nu = 1/2$ were made.  The complexity of those
measurements is greatly reduced if half-filling is retained while the
density is changed, as shown in Fig.\ 3, eliminating the effective
field.  A change in the peak position with density is still evident,
however the large variations in magnitude of the measurements of Fig.\
2 are now absent, with all densities approaching a common $\rho_D/T^2$
at higher temperatures.  The peak positions, shown in the inset, are
reasonably described by $T_P \propto \sqrt{n}$ (solid
line), consistent with changes of the size of the CF Fermi
surface.  This behavior does not result from a simple combination of
the individual dependences on field and density observed in Fig.\ 2.

\begin{figure}[tb]
\begin{center}
\leavevmode
\hbox{%
\epsfxsize=3in
\epsffile{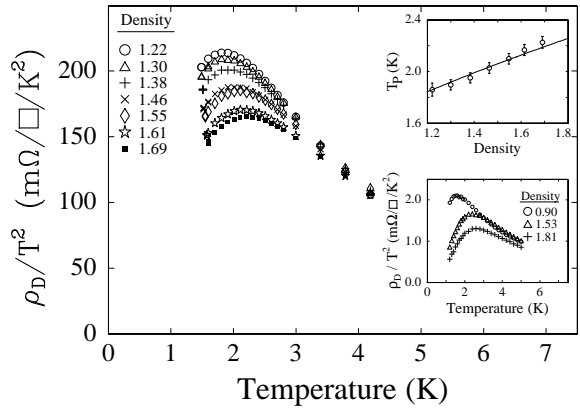}}
\end{center}
\caption{ Temperature dependence of $\rho_{D}/T^2$ for various carrier
densities showing similar behavior for composite Fermions and
zero-field electrons (lower inset).  For composite Fermions the field
was adjusted to maintain the system at $\nu=1/2$.  Upper Inset:
Changes in peak position, $T_P$, with density compared to $\sqrt{n}$
(solid line), which reflects the change in size of the CF Fermi
surface.}
\label{3}
\end{figure}

Comparison of the density dependence of the magnitude of $\rho_D$ in
Fig.\ 3 with that of electrons at zero field provides additional
support that this data results from a Fermi surface related cutoff of
CF scattering.  The electronic response, shown in the inset of Fig.\ 3
for the 2400 \AA\ barrier sample, reflects the general behavior of the
CF system.  In addition to $T_P$ varying with the size of the Fermi
surface, both show little density dependence in $\rho_D/T^2$ at high
temperatures despite the density of the electron measurements spanning
a wider range than for CF.  The striking similarity of the zero field
data with the CF measurements, when restricted to $\nu=1/2$, suggests
a simpler response in which the CF system mimics that of electrons.

These data raise a number of puzzling questions.  The first regards
behavior as $\nu$ is varied from half-filling.  The generally accepted
picture of an effective field which has little impact until the CF
cyclotron period is less than the scattering time is inconsistent with
the considerable changes observed in the density and field dependence
of $\rho_D$.  The origin of these inconsistencies and whether they are
related to the large q scattering probed in this work remains an open
question.

Another clearly important question involves the position of $T_P$
observed at half filling; it is one-third lower in temperature than
anticipated from extrapolation of the zero field measurements.
Various reasons for this discrepancy may be considered.  One possible
cause lies in the significant difference in sound velocity between
longitudinal and transverse phonons in GaAs layers.  The shift of
$T_P$ observed, however, would require that zero-field electrons
interact exclusively with longitudinal phonons, but CF predominantly
with transverse phonons.  Such behavior is inconsistent with both
theoretical investigations\cite{all:phdrag} of phonon drag and the
measured position of $T_P$ in the electron system.  A second
consideration is that the relative contribution of 2k$_F$ scattering,
as compared to smaller q's, may be substantially weaker for CF than in
the electron system.  Reducing this contribution could move $T_P$ to
lower temperatures.  This would contradict preliminary numerical
calculations\cite{rezayi:private} done for low energies.  Another
possibility is that the internal structure of the particles themselves
are probed in these large wavevector scattering events.  Resolution of
these and other questions raised in this work will likely require
additional investigation.

In summary, large wavevector scattering of composite Fermions has been
investigated through measurements of interlayer phonon drag.  The
temperature dependence of these measurements implies the existence of a
wavevector cutoff, in agreement with qualitative properties of the electron
system at zero field.  As the CF system is varied from $\nu = 1/2$, clear
changes in magnitude and temperature dependence develop which are
inconsistent with current expectations of CF's.  Varying the density but
remaining at half filling shows behavior substantially more consistent with
the zero field electron system.  A clear deviation remains, however, from
the temperature dependence anticipated for a wavevector cutoff
corresponding to $2k_F$ scattering.

\end{document}